\documentclass[%
 aip,
rsi,%
 amsmath,amssymb,
 reprint,%
]{revtex4-1}

\usepackage{amsmath}
\usepackage{array}
\usepackage{graphicx}% Include figure files
\usepackage{dcolumn}% Align table columns on decimal point
\usepackage{threeparttable}
\usepackage{bm}% bold math
\usepackage[colorlinks,linkcolor=blue,anchorcolor=blue,citecolor=blue,urlcolor=blue]{hyperref}
\usepackage{graphicx}
\begin{document}

\preprint{AIP/123-QED}

\title[Sample title]{Physical origin of the expansion of polymer coils in a binary solvent in the  vicinity of its demixing critical point}% Force line breaks with \\
%\thanks{Footnote to title of article.}

\author{Jan V. Sengers}
 \email{sengers@umd.edu}
 \affiliation{
Institute for Physical Science and Technology and Department of Chemical and Biomolecular Engineering, University of Maryland, College Park, MD 20742, USA}

\author{Mikhail A. Anisimov}
 \affiliation{
Institute for Physical Science and Technology and Department of Chemical and Biomolecular Engineering, University of Maryland, College Park, MD 20742, USA}

\author{Xiong Zheng}
 \affiliation{
Institute for Physical Science and Technology and Department of Chemical and Biomolecular Engineering, University of Maryland, College Park, MD 20742, USA}
\affiliation{
Key Laboratory of Thermo-Fluid Science and Engineering, Ministry of Education, Xi’an Jiaotong University, Xi’an, Shaanxi Province, 710049, P.R. China}

\date{\today}% It is always \today, today,

\begin{abstract}
Critical fluctuations are known to induce a collapse of polymer chains in a mixed solvent upon approaching its liquid-liquid critical point, as originally predicted by Brochard and de Gennes. Recently, we have found that closer to the critical point this collapse is followed by a reswelling of the polymer coils well beyond the original dimensions, a phenomenon not predicted by the theory of Brochard and de Gennes. We submit that upon approaching the critical temperature more closely, the correlation length of the critical fluctuations inside the polymer coils can no longer further increase due to the finite size of the coils, resulting in the appearance of large critical Casimir forces that cause a significant expansion of the polymer coils. Eventually, micro-phase separation inside the coils will appear and the coils will reshrink. This entire process takes place while the bulk solution is still in the one-phase region. \end{abstract}

%\keywords{critical fluctuations, liquid-liquid phase separation, polymer coil transformations, polymer solutions. }%Use showkeys class option if keyword
\maketitle

\section{\label{sec:level1}INTRODUCTION}

Critical fluctuations are known to cause a collapse of polymer chains upon approaching the critical point of a solvent. Specifically, Brochard and de Gennes predicted an initial collapse of the polymer chains followed by a reswelling of the polymer dimensions at the critical point to their original dimension \cite{de1976conformation,brochard1980collapse}. Support for this collapse and reswelling phenomenon has been provided by a large number of authors from alternative theoretical considerations \cite{vilgis1993conformation, vasilevskaya1998conformation, dua1999polymer, sumi2009critical}, computer simulations \cite{sumi2009critical, magda1988dimensions, vasilevskaya19981conformation, luna1997polymer, sumi2005anomalous, sumi2005cooperative, sumi2007behavior}, and experiments \cite{to1998polymer, morita2002polymer, grabowski2007contraction, he2012partial}.\par

\begin{figure*}[htbp]
\centering
\includegraphics[width=4.25in,height=2.7in]{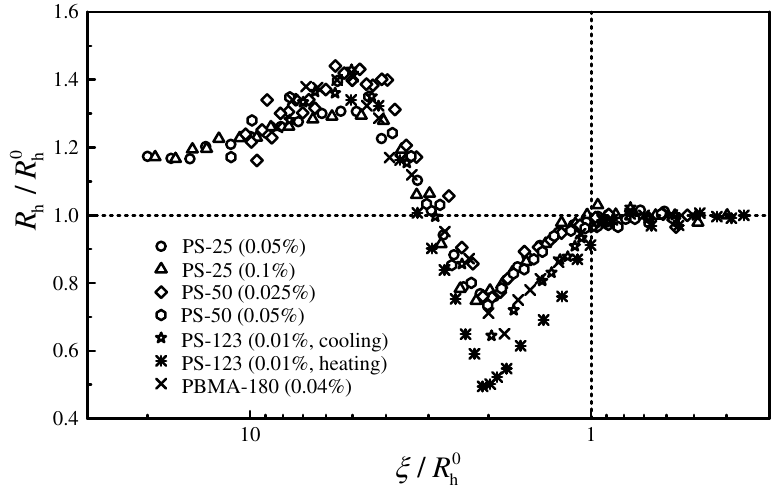}
\caption{\label{fig:fig1}The hydrodynamic radius \(R_\textup{h}\) of polymer chains as a function of the correlation length \(\xi\) of the critical fluctuations in the bulk solution for several polymer solutions, scaled in terms of the hydrodynamic radius \(R_\textup{h}^0\) of the polymer chains far away from the critical point. Abbreviations: PS-25, PS-50, PS-123, and PBMA-180 correspond to polystyrene with molecular weight \(M_\textup{w}\) = 25,000, 50,000, 123,000, and polybutylmethacrylate with \(M_\textup{w}\)  = 180,000. (Reproduced from Ref.~\onlinecite{zheng2018unusual}.)}
\end{figure*}

Recently, we studied the behavior of dilute polymer chains (polystyrene and poly-butyl methacrylate) of various molecular weights near the liquid-liquid critical point of a binary solvent (nitroethane+isooctane) by dynamic light scattering \cite{zheng2018unusual}. The mixture nitroethane+isooctane is special in that the difference between the refractive indices of the two liquid components is exceptionally small, so that the critical opalescence is weak while still measurable \cite{aref1973mandel, beysens1979coexistence}. Hence, we have been able to simultaneously observe the Brownian motion of the polymers and the diverging correlation length of the critical fluctuations closer to the critical temperature than previous investigators. To our surprise we found that the polymer coils exhibit a sequence of collapse-reswelling-expansion-reshrinking transitions upon approaching the critical temperature in the macroscopically homogeneous one-phase region of the polymer solution. The experimental results are summarized in Fig. \ref{fig:fig1} showing the hydrodynamic radius \(R_\textup{h}\) of the polymer coils as a function of the correlation length \(\xi\) of the critical fluctuations in the bulk solution, scaled in terms of the hydrodynamic radius \(R_\textup{h}^0\) of the polymer coils far away from the critical temperature \cite{zheng2018unusual}. The polymer coils begin to collapse when \(\xi/R_\textup{h}^0 \approx 1\), reach a minimum size when \(\xi/R_\textup{h}^0 \approx 2\), and then expand to a maximum value \(R_\textup{h}\) much larger than the original unperturbed value \(R_\textup{h}^0\) and finally shrinking again, a phenomenon not predicted by Brochard and de Gennes \cite{brochard1980collapse}. In retrospect, the experiments of Grabowski and Mukhopadhya \cite{grabowski2007contraction} and of He \textit{et al}. \cite{he2012partial} also showed already a trend of the coils to expand well beyond the unperturbed size \(R_\textup{h}^0\) when the critical temperature was approached more closely.\par

The value of the excluded volume controls the dimensions of polymer coils in the critical region \cite{flory1953principles}. As pointed out by Brochard and de Gennes \cite{brochard1980collapse}, the growing correlation length decreases the effective excluded volume since preferential adsorption of the better solvent forms a cloud near the polymer segments. This cloud attracts the neighboring segments and the polymer chain collapses. This collapse causes an increase of the monomer concentration inside the coils. Brochard and de Gennes assume that this increase of monomer concentration moves the system inside the coils away from the critical point, reducing the correlation length of the critical fluctuations, so that the coils reswell to their original size. Alternative theoretical scenarios for a collapse and subsequent reswelling to the original size have been proposed by Vilgis \textit{et al}. \cite{vilgis1993conformation}, Vasilevskaya \textit{et al}. \cite{vasilevskaya19981conformation}, and Dua and Cherayil \cite{dua1999polymer}. Particular interesting is the suggestion of Suma \textit{et al}. \cite{sumi2009critical} that the initial collapse is caused by critical Casimir forces. The idea is that critical Casimir forces may induce long-range attractive interactions between the polymers causing the initial collapse \cite{Schlesener2003Critical}. For instance, critical Casimir forces can be responsible for aggregation in colloids \cite{beysens1999wetting, Bonn2009Direct,Mohry2012Structure,guo2018nanoparticle,maciolek2018collective}. However, this picture cannot explain the subsequent experimentally reswelling phenomenon that we have observed.\par

This paper is organized as follows. In Section 2 we explain that upon approaching the critical temperature the solution inside the polymer coils enters the two-phase region while the bulk solution is still in the one-phase region. In Section 3 we propose that the experimentally observed expansion phenomenon may be caused by the appearance of critical Casimir forces inside the coils before the system inside the coils enters the two-phase region. Our conclusion is summarized in Section 4.\par

\section{Observed expansion of the polymer coils followed by microphase-separation}

A major problem with the current picture of the reswelling of polymer coils is the assumption of Brochard and de Gennes \cite{brochard1980collapse} that an increase of the monomer concentration due to the initial collapse causes the system inside the coil to move away from the critical point in the one-phase region. If the polymer would be equally soluble in both mixture components, the critical temperature would decrease with increasing polymer concentration in the case of a solution with an upper critical point and would increase with increasing polymer concentration in a solution with a lower critical point, thus in both cases shrinking the two-phase domain. However, the appearance of partial demixing in the solvent mixture indicates the presence of a significant difference in the molecular interactions of the mixture components, so that one-component is usually a significantly better solvent for the polymer than the other one. Nitroethane+isooctane has an upper critical temperature. Our experiments have clearly shown that addition of polymer does not decrease the critical temperature as assumed by Brochard and de Gennes, but instead increases the critical temperature, thus increasing the two-phase domain. Specifically, we found that, in the investigated range of polymer concentrations \textit{c}, the difference between the critical temperature \(T_\textup{c}(c)\) at polymer concentration \textit{c} and the critical temperature \(T_\textup{c}(0)\)  of the pure binary solvent, when scaled by the degree of polymerization \textit{N}, shows a universal linear dependence on \textit{c} with a positive coefficient \cite{zheng2018unusual}. In fact, addition of polymer causing an increase of the two-phase domain, appears to be a general phenomenon in phase separating polymer solutions \cite{to1998polymer, morita2002polymer, grabowski2007contraction, he2012partial}. The increase of the monomer concentration inside the coils, due to the initial collapse, causes the system inside the coils not to move away from its critical temperature as commonly has been assumed, but instead the system inside the coils will move closer to its critical temperature and will enter the two-phase region at a temperature where the bulk solution is still in the macroscopically homogeneous one-phase region. \par

\begin{figure}[htbp]
\centering
\includegraphics[width=3.25in,height=2.8in]{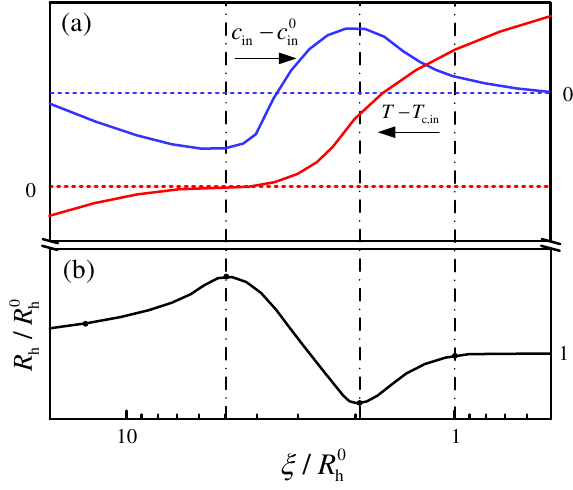}
\caption{\label{fig:fig2}(a) Schematic representation of the diﬀerence \(c_\textup{in}-c_\textup{in}^0\) between the monomer concentration (\(c_\textup{in}\)) inside the coil and the concentration (\(c_\textup{in}^0\)) far away from the critical temperature (blue curve) and the temperature diﬀerence \(T-T_\textup{c,in}\) (red curve) inside the coil as a function of the reduced correlation length \(\xi/R_\textup{h}^0\) measured in the bulk solution. (b) Schematic representation of the reduced hydrodynamic radius \(R_\textup{h}/R_\textup{h}^0\). The initial collapse of the hydrodynamic radius \(R_\textup{h}\) is accompanied by increasingly larger finite-size effects on the critical fluctuations, since the actual correlation length inside the coil cannot increase beyond the size of the polymer coil. The resulting critical Casimir forces of the system inside the coil become eventually large enough to cause an expansion of the coils. Finally, when \(T-T_\textup{c,in}\) goes through zero at a temperature where the solution is still above the critical temperature \(T_\textup{c}\), micro-phase separation occurs inside the coils and the coils will reshrink again.}
\end{figure}

To elucidate this behavior we show in Fig. \ref{fig:fig2} schematically the difference \(c_\textup{in}-c_\textup{in}^0\) between the actual polymer concentration \(c_\textup{in}\) inside the coil and the polymer concentration \(c_\textup{in}^0\) in the undisturbed coil far away from the critical point (blue curve), the difference \(T-T_\textup{c,in}\)  between the actual temperature \textit{T} and the critical temperature \(T_\textup{c,in}\) inside the coil (red curve), and the reduced hydrodynamic radius \(R_\textup{h}/R_\textup{h}^0\) as a function of the reduced correlation length \(\xi/R_\textup{h}^0\) measured in the bulk solution. \par

We want to emphasize the difference between the concentration of polymer in the bulk solution and the concentration of polymer monomers inside the coils. In fact, since the solution is very dilute, the bulk polymer concentration is not relevant for the transformations of individual polymer coils. However, the concentration of monomers inside the coil is significantly higher than the concentration in the bulk dilute solution. The concentration in terms of  molecular fraction of monomers inside the undisturbed coil, where \(R_\textup{h}^2 \propto N\) \cite{zheng2018unusual}, can be esimated as \(c_\textup{in}^0 \approx 1/N^{1/2}\) \cite{de1979scaling, Teraoka2002Polymer}. For \(N \approx 10^3 ( M \approx 10^5\)) \(c_\textup{in}^0 \approx 3 \cdot 10^{-2}\). Since \(T_\textup{c}(c)-T_\textup{c}(c=0) \approx c(dT_\textup{c}/dc)\) \cite{zheng2018unusual}, the system inside the coil will enter the two-phase region at \(T-T_\textup{c}=T_\textup{c,in}-T_\textup{c}=c_\textup{in}(dT_\textup{c}/dc_\textup{in}) \approx 0.1 ^\circ\)C assuming \(dT_\textup{c}/dc_\textup{in} \approx 0.3^\circ\)C. Unlike the critical tempearure of bulk polymer solutions where \(dT_\textup{c}/dc \approx 0.0015N^\circ\)C \cite{zheng2018unusual}, \(dT_\textup{c}/dc_\textup{in}\) should not depend on the degree of polymerization and is determined only by interactions of monomers with solvent molecules. However, the dependence \(T-T_\textup{c,in}\)  \textit{vs}. \(T-T_\textup{c}\) is not linear since \(c_\textup{in}\) increases and decreases when the coil shrinks and expands.\par

Upon approaching the critical temperature the polymer chains start to collapse due to a negative term in the excluded volume resulting from the increase of the correlation length, as predicted by Brochard and de Gennes \cite{brochard1980collapse}. The effect of reducing the hydrodynamic radius is significant, order of 30 \% (see Fig. \ref{fig:fig1}). This collapse causes an increase of the monomer concentration \(c_\textup{in}\) inside the coils, so that the critical temperature \(T_\textup{c,in}\) of the solution inside the coils, which is always larger than the critical temperature \(T_\textup{c}\) of the bulk solution, is moving faster to a higher value (see Fig. \ref{fig:fig2}). In principle, the preferential adsorption will also increase the nitroethane concentration inside the coils, but due to the ﬂatness of the coexistence curve its effect on the critical temperature is much smaller than that from the increase of the polymer concentration. Hence, as a result of the shrinking of the polymer coils, \(T-T_\textup{c,in}\) (inside the polymer coils) becomes increasingly smaller than \(T-T_\textup{c}\) of the bulk solution (not larger as assumed by Brochard and de Gennes!), thus causing an increase of the correlation length \(\xi\). \par

The collapse of the polymer coil is accompanied by the appearance of critical Casimir forces since the correlation length inside the coil cannot grow beyond the coil size (order of the radius of gyration \(R_\textup{g} \approx (3/2) R_\textup{h}\) \cite{Teraoka2002Polymer}). Eventually, these critical Casimir forces become so strong as to cause an expansion of the coils as further discussed in Section 3. The concentration inside the coil decreases, but \(T_\textup{c,in}\) is always larger than \(T_\textup{c}\) of the bulk solution, because \(c_\textup{in}\) is still significantly higher than the bulk concentration. Finally, the system inside the coil reaches a temperature \(T=T_\textup{c,in}\), while the bulk solution has not yet reached its critical temperature \(T_\textup{c} > T_\textup{c,in}\), and the system inside the coil will enter the two-phase region. Then micro-phase separation will occur inside the coil: one micro-phase enriched with nitroethane (good solvent) and the other micro-phase enriched with isooctane (poor solvent). The solution inside the coil will depart from the critical condition and the coil will reshrink to its original size. The experimental results, presented in Fig. \ref{fig:fig1}, have shown that the phenomenon is reversible, \textit{i.e.}, completely reproducible upon either decrease or increase of the temperature of the polymer solution \cite{zheng2018unusual}. Our interpretation is supported by a comparison of the temperatures of maximum expansion, \(T_\textup{max}\), observed for different degrees of polymerization (as seen in Fig. \ref{fig:fig3} below). Specifically, it appears that \(T_\textup{max}(N)-T_\textup{c} \propto c_\textup{in} \propto N^{-1/2}\), as we expected, namely, \(\sim 0.3^\circ\)C for PS-25, ~0.2\(^\circ\)C for PS-50, and ~0.1\(^\circ\)C for PS-123.\par

\begin{figure*}[htbp] 
\centering
\includegraphics[width=6.2in,height=5.1in]{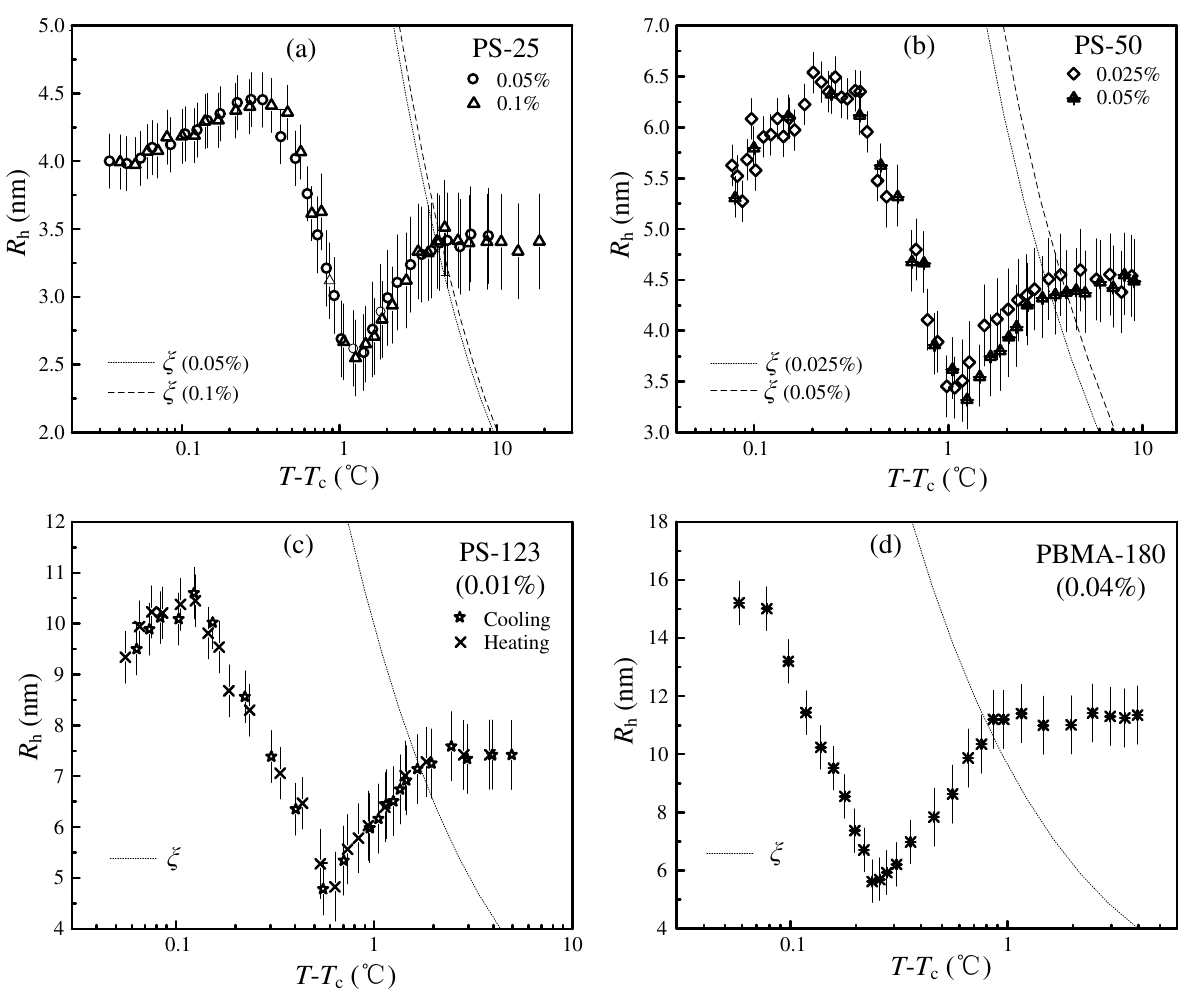}
\caption{\label{fig:fig3}The hydrodynamic radius \(R_\textup{h}\) of the polymer coils and the thermodynamic correlation length \(\xi\) in the bulk solution has a function of the distance of the temperature \textit{T} from the critical temperature \(T_\textup{c}\) of the bulk solution. (Reproduced from Ref.~\onlinecite{zheng2018unusual}.).}
\end{figure*}

One important issue that needs to be addressed is whether effects of co-solvency could explain the newly observed coil transformation. Indeed, the polymer coil size can be quite different not only in different solvents, but also the coil size can change in a mixed solvent \cite{magda1988dimensions}. To address this issue we have performed additional measurements of the coil sizes in pure isooctane and nitroethane, as well as in their mixtures, away from the critical temperature and in the two-phase region. \par

The hydrodynamic radius \(R_\textup{h}\) and the power-law exponent \textit{m} in the scaling relation \(R_\textup{h} \propto (M_\textup{w})^m\) for the increase of the hydrodynamic radius with increase of the molecular weight for the PS coils at different conditions are shown in Table ~\ref{table:radius}. Isooctane cannot dissolve PS and is a poor solvent for PS, while it is relatively close to a theta-solvent for PBMA. Nitroethane seems to be somewhere between a poor-solvent and a theta-solvent for both PS and PBMA. Nevertheless, PS in a mixed solvent is almost ideal and swells better than in the pure liquid components, which is attributed to the effect of co-solvency caused by unfavorable nitroethane-isooctane interactions \cite{Sander1986Solubility}. The exponent \textit{m} increases significantly from \textit{m} = 0.39 for the collapsed chain to \textit{m} = 0.55 for the fully expanded chain. Hence, the data in Table ~\ref{table:radius} confirm that the effects of shrinking and expansion, which we have observed close to the critical point, are much more significant than the variations in the sizes observed in different solvents, as well as in the mixed solvent away from the critical temperature.\par

\begin{table} 
\centering
\caption{\label{table:radius}Hydrodynamic radius  (\(R_\textup{h}\)/nm) and the power-law exponent \textit{m} in molecular-weight scaling of polystyrene coils}
\begin{threeparttable}
\begin{tabular}{lcccc}
   \hline\hline
    & PS-25 & PS-50 & PS-123 & \(\textit{m}\)\\
   \hline
Nitroethane\tnote{1}  & 3.11\(\pm0.30\) & 4.09\(\pm0.40\) & 6.15\(\pm0.60\) & 0.47\(\pm0.01\)\\
Mixed solvent\tnote{1}  & 3.40\(\pm0.35\) & 4.54\(\pm0.40\) & 7.42\(\pm0.68\) & 0.49\(\pm0.01\) \\
Lower phase\tnote{2} & / & / & 6.45\(\pm0.60\) & /\\
Collapsed & 2.55\(\pm0.28\) & 3.40\(\pm0.30\) & 4.78\(\pm0.51\) & 0.39\(\pm0.01\)\\
Expanded & 4.40\(\pm0.20\) & 6.54\(\pm0.20\) & 10.6\(\pm0.51\) & 0.55\(\pm0.01\)\\
   \hline\hline
\end{tabular}
\begin{tablenotes}
        \footnotesize
        \item[1] 50 $^{\circ}$C
        \item[2] in the two-phase region at 25 $^{\circ}$C; the lower phase contains mainly nitroethane
      \end{tablenotes}
    \end{threeparttable}
\end{table}

We want also to clarify the relationship between hydrodynamic radius and radius of gyration of polymer chains. For a linear polymer chain, the hydrodynamic radius is proportional to the radius of gyration. The value of \(R_\textup{g}/R_\textup{h}\) ranges from 1.5 to 1.6 when the solvent changes from a good to a theta solvent. Nitroethane+isooctane is a better solvent than a theta solvent for polystyrene, so that \(R_\textup{g}/R_\textup{h}\) is between 1.5 and 1.6 for the polymers in nitroethane+isooctane. The experimental results show that the collapse begins when \(\xi \approx R_\textup{h}\); it is expected that the collapse begins when \(\xi \approx 0.6R_\textup{g}\). This observation is in full agreement with the prediction of Brochard and de Gennes.\par

\section{Role of critical Casimir forces}

Long-range critical fluctuations will cause an effective force in confined liquid layers near a critical point as originally predicted by Fisher and de Gennes \cite{Fisher1978Wall}. The phenomenon has been studied extensively both theoretically and experimentally \cite{Krech1994The, krech1997casimir, krech1999fluctuation, hertlein2008gambassi, gambassi2009critical}. Critical Casimir forces can be either attractive or repulsive depending on the boundary conditions \cite{Krech1994The, krech1997casimir, krech1999fluctuation, rafai2007repulsive}, \textit{i.e.}, attractive for a symmetric system (confined between two equally solvophilic or solvophobic boundaries) or repulsive for an asymmetric system (one boundary being solvophilic and the other one being solvophobic).\par

When a critical fluid is confined between two surfaces with an area \textit{A} and separated by a distance \textit{L}, the fluid will exert a Casimir force F on the surfaces, given by \cite{gambassi2009critical}

\begin{equation}
F=\frac{k_\textup{B}TA}{L^3}\Theta(L/\xi)\label{con:1},
\end{equation}

\noindent where \(k_\textup{B}\) is Boltzmann’s constant, \textit{T} the temperature, and where \(\Theta(L/\xi)\) is a finite-size scaling function. Near the critical temperature \(T_\textup{c}\) the correlation length \(\xi\) diverges as

\begin{equation}
\xi=\xi_\textup{0}(\Delta T/T_\textup{c})^{-\nu},\label{con:2}
\end{equation}

\noindent where \(\xi_\textup{0}\) is an amplitude of the order of the molecular interaction range and where \(\nu=0.63\) is a universal critical exponent \cite{sengers2009experimental}. Eq. \ref{con:1} is valid when \(\xi_\textup{0} \ll L \ll \xi\) \cite{gambassi2009critical}. One commonly defines a universal critical amplitude \(\Theta=$${\lim_{y\to0}}$$ \Theta(y)\), whose sign and magnitude, however, depends on the boundary conditions. For the three-dimensional Ising universality class with asymmetric boundary conditions (+-), the experimental value is \(\Theta_\textup{\(+-\)}=+6\pm2\) \cite{Fukuto2005Critical} in agreement with theoretical estimates \cite{vasilyev2009universal, hasenbusch2010thermodynamic}. As explained in the previous section when the size of the polymer coil begins to decrease, the correlation length \(\xi\) inside the coil can no longer grow because of the finite size of the polymer coil. Correspondingly, the osmotic susceptibility \(\chi\) of the system inside the coil can no longer diverge as explained in our previous publication \cite{zheng2018unusual}. These are exactly the type of boundary effects causing the appearance of critical Casimir forces.\par

To illustrate the importance of critical Casimir forces in our experimental observation of the expansion of the polymer coils close to the critical point of the polymer solution, we show in Fig. \ref{fig:fig3} both the hydrodynamic radius \(R_\textup{h}\) of the various polystyrene samples as well as the correlation length \(\xi\) in the bulk solution as a function of \(T-T_\textup{c}\), where \(T_\textup{c}\) is the critical temperature of the bulk solution. We see that at the temperature where \(R_\textup{h}\) has reached its smallest value, the correlation length \(\xi\) in the bulk solution is already a magnitude larger than the size of the polymer coils. Moreover, as pointed out earlier, \(T-T_\textup{c,in}\) inside the polymer coils is even smaller than \(T-T_\textup{c}\), so that the correlation length \(\xi\) in the thermodynamic limit corresponding to the state of the system inside the coils would even be another magnitude larger than the size of the polymer coils. However, the actual correlation length inside the polymer coils cannot become larger than the size of the polymer coils thus inducing huge critical Casimir pressures. Our experiments show that these critical Casimir forces, arising because the correlation length inside the coils wants to increase, are clearly repulsive, as one would expect intuitively. For a quantitative analysis we need an explicit expression for critical Casimir forces inside a single (spherical) boundary at temperatures, where the corresponding correlation length becomes order of magnitude larger than the size of the spherical volume, which to our knowledge is not currently available. If we naively substitute into Eq. \ref{con:1}, which is an example of an expression for repulsive critical Casimir pressures,  \(L \approx R_\textup{h}\), when \(R_\textup{h}\) reaches its minimum value (varying from 2.5 nm to 5.0 nm depending on the polymer molecular weight of our polystyrene samples), the critical Casimir pressures would even exceed macroscopic values of the order of 0.1 MPa.\par
 
 The bottom line is that our experiments have revealed that at the temperatures where finite-size effects on the critical fluctuations (inside the polymer coils) become enormously large, thus inducing huge critical Casimir forces inside the coils, we see a pronounced expansion of the polymer coils.\par
 
 \section{Conclusion}

The initial collapse of the polymer coils, predicted by Brochard and de Gennes \cite{brochard1980collapse}, will continue not until the system inside the coils is moving away from its critical temperature but until the critical Casimir forces induced by the finite-size effects on the critical fluctuations inside the coils become so large that they cause an expansion of the polymer coils. This expansion will continue till eventually the system inside the coils enters two-phase region where micro-phase will appear, while the bulk solution is still in a homogeneous state.\par

\begin{acknowledgments}
The authors acknowledge stimulating discussion with Siegfried Dietrich of the Max-Planck Institute for Intelligent Systems and Deverajan Thirumalai of the University of Texas. The enthusiastic support of Maogang He of Xi’an Jiaotong University for this project is also gratefully acknowledged. The research was supported by Grant No. 59434-ND6 of the Petroleum Research Fund of the American Chemical Society. Xiong Zheng was supported by Grant No. 201606280242 of the China Scholarship Council.
\end{acknowledgments}

%\nocite{*}
\bibliography{main}% Produces the bibliography via BibTeX.

\begin{thebibliography}{40}%
\makeatletter
\providecommand \@ifxundefined [1]{%
 \@ifx{#1\undefined}
}%
\providecommand \@ifnum [1]{%
 \ifnum #1\expandafter \@firstoftwo
 \else \expandafter \@secondoftwo
 \fi
}%
\providecommand \@ifx [1]{%
 \ifx #1\expandafter \@firstoftwo
 \else \expandafter \@secondoftwo
 \fi
}%
\providecommand \natexlab [1]{#1}%
\providecommand \enquote  [1]{``#1''}%
\providecommand \bibnamefont  [1]{#1}%
\providecommand \bibfnamefont [1]{#1}%
\providecommand \citenamefont [1]{#1}%
\providecommand \href@noop [0]{\@secondoftwo}%
\providecommand \href [0]{\begingroup \@sanitize@url \@href}%
\providecommand \@href[1]{\@@startlink{#1}\@@href}%
\providecommand \@@href[1]{\endgroup#1\@@endlink}%
\providecommand \@sanitize@url [0]{\catcode `\\12\catcode `\$12\catcode
  `\&12\catcode `\#12\catcode `\^12\catcode `\_12\catcode `\%12\relax}%
\providecommand \@@startlink[1]{}%
\providecommand \@@endlink[0]{}%
\providecommand \url  [0]{\begingroup\@sanitize@url \@url }%
\providecommand \@url [1]{\endgroup\@href {#1}{\urlprefix }}%
\providecommand \urlprefix  [0]{URL }%
\providecommand \Eprint [0]{\href }%
\providecommand \doibase [0]{http://dx.doi.org/}%
\providecommand \selectlanguage [0]{\@gobble}%
\providecommand \bibinfo  [0]{\@secondoftwo}%
\providecommand \bibfield  [0]{\@secondoftwo}%
\providecommand \translation [1]{[#1]}%
\providecommand \BibitemOpen [0]{}%
\providecommand \bibitemStop [0]{}%
\providecommand \bibitemNoStop [0]{.\EOS\space}%
\providecommand \EOS [0]{\spacefactor3000\relax}%
\providecommand \BibitemShut  [1]{\csname bibitem#1\endcsname}%
\let\auto@bib@innerbib\@empty
%</preamble>
\bibitem [{\citenamefont {De~Gennes}(1976)}]{de1976conformation}%
  \BibitemOpen
  \bibfield  {author} {\bibinfo {author} {\bibfnamefont {P.}~\bibnamefont
  {De~Gennes}},\ }\href@noop {} {\bibfield  {journal} {\bibinfo  {journal}
  {Journal de Physique Lettres}\ }\textbf {\bibinfo {volume} {37}},\ \bibinfo
  {pages} {59} (\bibinfo {year} {1976})}\BibitemShut {NoStop}%
\bibitem [{\citenamefont {Brochard}\ and\ \citenamefont
  {De~Gennes}(1980)}]{brochard1980collapse}%
  \BibitemOpen
  \bibfield  {author} {\bibinfo {author} {\bibfnamefont {F.}~\bibnamefont
  {Brochard}}\ and\ \bibinfo {author} {\bibfnamefont {P.~G.}\ \bibnamefont
  {De~Gennes}},\ }\href@noop {} {\bibfield  {journal} {\bibinfo  {journal}
  {Ferroelectrics}\ }\textbf {\bibinfo {volume} {30}},\ \bibinfo {pages} {33}
  (\bibinfo {year} {1980})}\BibitemShut {NoStop}%
\bibitem [{\citenamefont {Vilgis}, \citenamefont {Sans},\ and\ \citenamefont
  {Jannink}(1993)}]{vilgis1993conformation}%
  \BibitemOpen
  \bibfield  {author} {\bibinfo {author} {\bibfnamefont {T.}~\bibnamefont
  {Vilgis}}, \bibinfo {author} {\bibfnamefont {A.}~\bibnamefont {Sans}}, \ and\
  \bibinfo {author} {\bibfnamefont {G.}~\bibnamefont {Jannink}},\ }\href@noop
  {} {\bibfield  {journal} {\bibinfo  {journal} {Journal de Physique II}\
  }\textbf {\bibinfo {volume} {3}},\ \bibinfo {pages} {1779} (\bibinfo {year}
  {1993})}\BibitemShut {NoStop}%
\bibitem [{\citenamefont {Vasilevskaya}, \citenamefont {Khalatur},\ and\
  \citenamefont {Khokhlov}(1998{\natexlab{a}})}]{vasilevskaya1998conformation}%
  \BibitemOpen
  \bibfield  {author} {\bibinfo {author} {\bibfnamefont {V.~V.}\ \bibnamefont
  {Vasilevskaya}}, \bibinfo {author} {\bibfnamefont {P.~G.}\ \bibnamefont
  {Khalatur}}, \ and\ \bibinfo {author} {\bibfnamefont {A.~R.}\ \bibnamefont
  {Khokhlov}},\ }\href@noop {} {\bibfield  {journal} {\bibinfo  {journal} {The
  Journal of chemical physics}\ }\textbf {\bibinfo {volume} {109}},\ \bibinfo
  {pages} {5108} (\bibinfo {year} {1998}{\natexlab{a}})}\BibitemShut {NoStop}%
\bibitem [{\citenamefont {Dua}\ and\ \citenamefont
  {Cherayil}(1999)}]{dua1999polymer}%
  \BibitemOpen
  \bibfield  {author} {\bibinfo {author} {\bibfnamefont {A.}~\bibnamefont
  {Dua}}\ and\ \bibinfo {author} {\bibfnamefont {B.~J.}\ \bibnamefont
  {Cherayil}},\ }\href@noop {} {\bibfield  {journal} {\bibinfo  {journal} {The
  Journal of chemical physics}\ }\textbf {\bibinfo {volume} {111}},\ \bibinfo
  {pages} {3274} (\bibinfo {year} {1999})}\BibitemShut {NoStop}%
\bibitem [{\citenamefont {Sumi}, \citenamefont {Imazaki},\ and\ \citenamefont
  {Sekino}(2009)}]{sumi2009critical}%
  \BibitemOpen
  \bibfield  {author} {\bibinfo {author} {\bibfnamefont {T.}~\bibnamefont
  {Sumi}}, \bibinfo {author} {\bibfnamefont {N.}~\bibnamefont {Imazaki}}, \
  and\ \bibinfo {author} {\bibfnamefont {H.}~\bibnamefont {Sekino}},\
  }\href@noop {} {\bibfield  {journal} {\bibinfo  {journal} {Phys. Rev. E}\
  }\textbf {\bibinfo {volume} {79}},\ \bibinfo {pages} {030801} (\bibinfo
  {year} {2009})}\BibitemShut {NoStop}%
\bibitem [{\citenamefont {Magda}\ \emph {et~al.}(1988)\citenamefont {Magda},
  \citenamefont {Fredrickson}, \citenamefont {Larson},\ and\ \citenamefont
  {Helfand}}]{magda1988dimensions}%
  \BibitemOpen
  \bibfield  {author} {\bibinfo {author} {\bibfnamefont {J.~J.}\ \bibnamefont
  {Magda}}, \bibinfo {author} {\bibfnamefont {G.~H.}\ \bibnamefont
  {Fredrickson}}, \bibinfo {author} {\bibfnamefont {R.~G.}\ \bibnamefont
  {Larson}}, \ and\ \bibinfo {author} {\bibfnamefont {E.}~\bibnamefont
  {Helfand}},\ }\href@noop {} {\bibfield  {journal} {\bibinfo  {journal}
  {Macromolecules}\ }\textbf {\bibinfo {volume} {21}},\ \bibinfo {pages} {726}
  (\bibinfo {year} {1988})}\BibitemShut {NoStop}%
\bibitem [{\citenamefont {Vasilevskaya}, \citenamefont {Khalatur},\ and\
  \citenamefont
  {Khokhlov}(1998{\natexlab{b}})}]{vasilevskaya19981conformation}%
  \BibitemOpen
  \bibfield  {author} {\bibinfo {author} {\bibfnamefont {V.~V.}\ \bibnamefont
  {Vasilevskaya}}, \bibinfo {author} {\bibfnamefont {P.~G.}\ \bibnamefont
  {Khalatur}}, \ and\ \bibinfo {author} {\bibfnamefont {A.~R.}\ \bibnamefont
  {Khokhlov}},\ }\href@noop {} {\bibfield  {journal} {\bibinfo  {journal} {The
  Journal of chemical physics}\ }\textbf {\bibinfo {volume} {109}},\ \bibinfo
  {pages} {5119} (\bibinfo {year} {1998}{\natexlab{b}})}\BibitemShut {NoStop}%
\bibitem [{\citenamefont {Luna-B{\'a}rcenas}\ \emph {et~al.}(1997)\citenamefont
  {Luna-B{\'a}rcenas}, \citenamefont {Gromov}, \citenamefont {Meredith},
  \citenamefont {Sanchez}, \citenamefont {de~Pablo},\ and\ \citenamefont
  {Johnston}}]{luna1997polymer}%
  \BibitemOpen
  \bibfield  {author} {\bibinfo {author} {\bibfnamefont {G.}~\bibnamefont
  {Luna-B{\'a}rcenas}}, \bibinfo {author} {\bibfnamefont {D.~G.}\ \bibnamefont
  {Gromov}}, \bibinfo {author} {\bibfnamefont {J.~C.}\ \bibnamefont
  {Meredith}}, \bibinfo {author} {\bibfnamefont {I.~C.}\ \bibnamefont
  {Sanchez}}, \bibinfo {author} {\bibfnamefont {J.~J.}\ \bibnamefont
  {de~Pablo}}, \ and\ \bibinfo {author} {\bibfnamefont {K.~P.}\ \bibnamefont
  {Johnston}},\ }\href@noop {} {\bibfield  {journal} {\bibinfo  {journal}
  {Chem. Phys. Lett.}\ }\textbf {\bibinfo {volume} {278}},\ \bibinfo {pages}
  {302} (\bibinfo {year} {1997})}\BibitemShut {NoStop}%
\bibitem [{\citenamefont {Sumi}\ and\ \citenamefont
  {Sekino}(2005{\natexlab{a}})}]{sumi2005anomalous}%
  \BibitemOpen
  \bibfield  {author} {\bibinfo {author} {\bibfnamefont {T.}~\bibnamefont
  {Sumi}}\ and\ \bibinfo {author} {\bibfnamefont {H.}~\bibnamefont {Sekino}},\
  }\href@noop {} {\bibfield  {journal} {\bibinfo  {journal} {Chem. Phys.
  Lett.}\ }\textbf {\bibinfo {volume} {407}},\ \bibinfo {pages} {322} (\bibinfo
  {year} {2005}{\natexlab{a}})}\BibitemShut {NoStop}%
\bibitem [{\citenamefont {Sumi}\ and\ \citenamefont
  {Sekino}(2005{\natexlab{b}})}]{sumi2005cooperative}%
  \BibitemOpen
  \bibfield  {author} {\bibinfo {author} {\bibfnamefont {T.}~\bibnamefont
  {Sumi}}\ and\ \bibinfo {author} {\bibfnamefont {H.}~\bibnamefont {Sekino}},\
  }\href@noop {} {\bibfield  {journal} {\bibinfo  {journal} {J. Chem. Phys.}\
  }\textbf {\bibinfo {volume} {122}},\ \bibinfo {pages} {194910} (\bibinfo
  {year} {2005}{\natexlab{b}})}\BibitemShut {NoStop}%
\bibitem [{\citenamefont {Sumi}, \citenamefont {Kobayashi},\ and\ \citenamefont
  {Sekino}(2007)}]{sumi2007behavior}%
  \BibitemOpen
  \bibfield  {author} {\bibinfo {author} {\bibfnamefont {T.}~\bibnamefont
  {Sumi}}, \bibinfo {author} {\bibfnamefont {K.}~\bibnamefont {Kobayashi}}, \
  and\ \bibinfo {author} {\bibfnamefont {H.}~\bibnamefont {Sekino}},\
  }\href@noop {} {\bibfield  {journal} {\bibinfo  {journal} {J. Chem. Phys.}\
  }\textbf {\bibinfo {volume} {127}},\ \bibinfo {pages} {164904} (\bibinfo
  {year} {2007})}\BibitemShut {NoStop}%
\bibitem [{\citenamefont {To}\ and\ \citenamefont
  {Choi}(1998)}]{to1998polymer}%
  \BibitemOpen
  \bibfield  {author} {\bibinfo {author} {\bibfnamefont {K.}~\bibnamefont
  {To}}\ and\ \bibinfo {author} {\bibfnamefont {H.~J.}\ \bibnamefont {Choi}},\
  }\href@noop {} {\bibfield  {journal} {\bibinfo  {journal} {Phys. Rev. Lett.}\
  }\textbf {\bibinfo {volume} {80}},\ \bibinfo {pages} {536} (\bibinfo {year}
  {1998})}\BibitemShut {NoStop}%
\bibitem [{\citenamefont {Morita}, \citenamefont {Tsunomori},\ and\
  \citenamefont {Ushiki}(2002)}]{morita2002polymer}%
  \BibitemOpen
  \bibfield  {author} {\bibinfo {author} {\bibfnamefont {S.}~\bibnamefont
  {Morita}}, \bibinfo {author} {\bibfnamefont {F.}~\bibnamefont {Tsunomori}}, \
  and\ \bibinfo {author} {\bibfnamefont {H.}~\bibnamefont {Ushiki}},\
  }\href@noop {} {\bibfield  {journal} {\bibinfo  {journal} {Eur. Polym. J.}\
  }\textbf {\bibinfo {volume} {38}},\ \bibinfo {pages} {1863} (\bibinfo {year}
  {2002})}\BibitemShut {NoStop}%
\bibitem [{\citenamefont {Grabowski}\ and\ \citenamefont
  {Mukhopadhyay}(2007)}]{grabowski2007contraction}%
  \BibitemOpen
  \bibfield  {author} {\bibinfo {author} {\bibfnamefont {C.~A.}\ \bibnamefont
  {Grabowski}}\ and\ \bibinfo {author} {\bibfnamefont {A.}~\bibnamefont
  {Mukhopadhyay}},\ }\href@noop {} {\bibfield  {journal} {\bibinfo  {journal}
  {Phys. Rev. Lett.}\ }\textbf {\bibinfo {volume} {98}},\ \bibinfo {pages}
  {207801} (\bibinfo {year} {2007})}\BibitemShut {NoStop}%
\bibitem [{\citenamefont {He}, \citenamefont {Cheng},\ and\ \citenamefont
  {Melnichenko}(2012)}]{he2012partial}%
  \BibitemOpen
  \bibfield  {author} {\bibinfo {author} {\bibfnamefont {L.}~\bibnamefont
  {He}}, \bibinfo {author} {\bibfnamefont {G.}~\bibnamefont {Cheng}}, \ and\
  \bibinfo {author} {\bibfnamefont {Y.~B.}\ \bibnamefont {Melnichenko}},\
  }\href@noop {} {\bibfield  {journal} {\bibinfo  {journal} {Phys. Rev. Lett.}\
  }\textbf {\bibinfo {volume} {109}},\ \bibinfo {pages} {067801} (\bibinfo
  {year} {2012})}\BibitemShut {NoStop}%
\bibitem [{\citenamefont {Zheng}\ \emph {et~al.}(2018)\citenamefont {Zheng},
  \citenamefont {Anisimov}, \citenamefont {Sengers},\ and\ \citenamefont
  {He}}]{zheng2018unusual}%
  \BibitemOpen
  \bibfield  {author} {\bibinfo {author} {\bibfnamefont {X.}~\bibnamefont
  {Zheng}}, \bibinfo {author} {\bibfnamefont {M.~A.}\ \bibnamefont {Anisimov}},
  \bibinfo {author} {\bibfnamefont {J.~V.}\ \bibnamefont {Sengers}}, \ and\
  \bibinfo {author} {\bibfnamefont {M.}~\bibnamefont {He}},\ }\href@noop {}
  {\bibfield  {journal} {\bibinfo  {journal} {Physical review letters}\
  }\textbf {\bibinfo {volume} {121}},\ \bibinfo {pages} {207802} (\bibinfo
  {year} {2018})}\BibitemShut {NoStop}%
\bibitem [{\citenamefont {Aref'ev}\ \emph {et~al.}(1973)\citenamefont
  {Aref'ev}, \citenamefont {Fabelinskii}, \citenamefont {Anisimov},
  \citenamefont {Kiyachenko},\ and\ \citenamefont {Voronov}}]{aref1973mandel}%
  \BibitemOpen
  \bibfield  {author} {\bibinfo {author} {\bibfnamefont {I.~M.}\ \bibnamefont
  {Aref'ev}}, \bibinfo {author} {\bibfnamefont {I.~L.}\ \bibnamefont
  {Fabelinskii}}, \bibinfo {author} {\bibfnamefont {M.~A.}\ \bibnamefont
  {Anisimov}}, \bibinfo {author} {\bibfnamefont {Y.~F.}\ \bibnamefont
  {Kiyachenko}}, \ and\ \bibinfo {author} {\bibfnamefont {V.~P.}\ \bibnamefont
  {Voronov}},\ }\href@noop {} {\bibfield  {journal} {\bibinfo  {journal} {Opt.
  Commun.}\ }\textbf {\bibinfo {volume} {9}},\ \bibinfo {pages} {69} (\bibinfo
  {year} {1973})}\BibitemShut {NoStop}%
\bibitem [{\citenamefont {Beysens}(1979)}]{beysens1979coexistence}%
  \BibitemOpen
  \bibfield  {author} {\bibinfo {author} {\bibfnamefont {D.}~\bibnamefont
  {Beysens}},\ }\href@noop {} {\bibfield  {journal} {\bibinfo  {journal} {J.
  Chem. Phys.}\ }\textbf {\bibinfo {volume} {71}},\ \bibinfo {pages} {2557}
  (\bibinfo {year} {1979})}\BibitemShut {NoStop}%
\bibitem [{\citenamefont {Flory}(1953)}]{flory1953principles}%
  \BibitemOpen
  \bibfield  {author} {\bibinfo {author} {\bibfnamefont {P.~J.}\ \bibnamefont
  {Flory}},\ }\href@noop {} {\emph {\bibinfo {title} {Principles of Polymer
  Chemistry}}}\ (\bibinfo  {publisher} {Cornell University Press, New York},\
  \bibinfo {year} {1953})\BibitemShut {NoStop}%
\bibitem [{\citenamefont {Schlesener}, \citenamefont {Hanke},\ and\
  \citenamefont {Dietrich}(2003)}]{Schlesener2003Critical}%
  \BibitemOpen
  \bibfield  {author} {\bibinfo {author} {\bibfnamefont {F.}~\bibnamefont
  {Schlesener}}, \bibinfo {author} {\bibfnamefont {A.}~\bibnamefont {Hanke}}, \
  and\ \bibinfo {author} {\bibfnamefont {S.}~\bibnamefont {Dietrich}},\
  }\href@noop {} {\bibfield  {journal} {\bibinfo  {journal} {Journal of
  statistical physics}\ }\textbf {\bibinfo {volume} {110}},\ \bibinfo {pages}
  {981} (\bibinfo {year} {2003})}\BibitemShut {NoStop}%
\bibitem [{\citenamefont {Beysens}\ and\ \citenamefont
  {Narayanan}(1999)}]{beysens1999wetting}%
  \BibitemOpen
  \bibfield  {author} {\bibinfo {author} {\bibfnamefont {D.}~\bibnamefont
  {Beysens}}\ and\ \bibinfo {author} {\bibfnamefont {T.}~\bibnamefont
  {Narayanan}},\ }\href@noop {} {\bibfield  {journal} {\bibinfo  {journal} {J.
  Stat. Phys.}\ }\textbf {\bibinfo {volume} {95}},\ \bibinfo {pages} {997}
  (\bibinfo {year} {1999})}\BibitemShut {NoStop}%
\bibitem [{\citenamefont {Bonn}\ \emph {et~al.}(2009)\citenamefont {Bonn},
  \citenamefont {Otwinowski}, \citenamefont {Sacanna}, \citenamefont {Guo},
  \citenamefont {Wegdam},\ and\ \citenamefont {Schall}}]{Bonn2009Direct}%
  \BibitemOpen
  \bibfield  {author} {\bibinfo {author} {\bibfnamefont {D.}~\bibnamefont
  {Bonn}}, \bibinfo {author} {\bibfnamefont {J.}~\bibnamefont {Otwinowski}},
  \bibinfo {author} {\bibfnamefont {S.}~\bibnamefont {Sacanna}}, \bibinfo
  {author} {\bibfnamefont {H.}~\bibnamefont {Guo}}, \bibinfo {author}
  {\bibfnamefont {G.}~\bibnamefont {Wegdam}}, \ and\ \bibinfo {author}
  {\bibfnamefont {P.}~\bibnamefont {Schall}},\ }\href@noop {} {\bibfield
  {journal} {\bibinfo  {journal} {Phys. Rev. Lett.}\ }\textbf {\bibinfo
  {volume} {103}},\ \bibinfo {pages} {156101} (\bibinfo {year}
  {2009})}\BibitemShut {NoStop}%
\bibitem [{\citenamefont {Mohry}, \citenamefont {Macio{\l}ek},\ and\
  \citenamefont {Dietrich}(2012)}]{Mohry2012Structure}%
  \BibitemOpen
  \bibfield  {author} {\bibinfo {author} {\bibfnamefont {T.~F.}\ \bibnamefont
  {Mohry}}, \bibinfo {author} {\bibfnamefont {A.}~\bibnamefont {Macio{\l}ek}},
  \ and\ \bibinfo {author} {\bibfnamefont {S.}~\bibnamefont {Dietrich}},\
  }\href@noop {} {\bibfield  {journal} {\bibinfo  {journal} {J. Chem. Phys.}\
  }\textbf {\bibinfo {volume} {136}},\ \bibinfo {pages} {224903} (\bibinfo
  {year} {2012})}\BibitemShut {NoStop}%
\bibitem [{\citenamefont {Guo}, \citenamefont {Stan},\ and\ \citenamefont
  {Liu}(2018)}]{guo2018nanoparticle}%
  \BibitemOpen
  \bibfield  {author} {\bibinfo {author} {\bibfnamefont {H.}~\bibnamefont
  {Guo}}, \bibinfo {author} {\bibfnamefont {G.}~\bibnamefont {Stan}}, \ and\
  \bibinfo {author} {\bibfnamefont {Y.}~\bibnamefont {Liu}},\ }\href@noop {}
  {\bibfield  {journal} {\bibinfo  {journal} {Soft matter}\ }\textbf {\bibinfo
  {volume} {14}},\ \bibinfo {pages} {1311} (\bibinfo {year}
  {2018})}\BibitemShut {NoStop}%
\bibitem [{\citenamefont {Macio{\l}ek}\ and\ \citenamefont
  {Dietrich}(2018)}]{maciolek2018collective}%
  \BibitemOpen
  \bibfield  {author} {\bibinfo {author} {\bibfnamefont {A.}~\bibnamefont
  {Macio{\l}ek}}\ and\ \bibinfo {author} {\bibfnamefont {S.}~\bibnamefont
  {Dietrich}},\ }\href@noop {} {\bibfield  {journal} {\bibinfo  {journal}
  {Reviews of Modern Physics}\ }\textbf {\bibinfo {volume} {90}},\ \bibinfo
  {pages} {045001} (\bibinfo {year} {2018})}\BibitemShut {NoStop}%
\bibitem [{\citenamefont {De~Gennes}\ and\ \citenamefont
  {Gennes}(1979)}]{de1979scaling}%
  \BibitemOpen
  \bibfield  {author} {\bibinfo {author} {\bibfnamefont {P.-G.}\ \bibnamefont
  {De~Gennes}}\ and\ \bibinfo {author} {\bibfnamefont {P.-G.}\ \bibnamefont
  {Gennes}},\ }\href@noop {} {\emph {\bibinfo {title} {Scaling Concepts in
  Polymer Physics}}}\ (\bibinfo  {publisher} {Cornell University Press, New
  York},\ \bibinfo {year} {1979})\BibitemShut {NoStop}%
\bibitem [{\citenamefont {Teraoka}(2002)}]{Teraoka2002Polymer}%
  \BibitemOpen
  \bibfield  {author} {\bibinfo {author} {\bibfnamefont {I.}~\bibnamefont
  {Teraoka}},\ }\href@noop {} {\emph {\bibinfo {title} {Polymer Solutions: An
  Introduction to Physical Properties}}}\ (\bibinfo  {publisher} {Wiley, New
  York},\ \bibinfo {year} {2002})\BibitemShut {NoStop}%
\bibitem [{\citenamefont {Sander}\ and\ \citenamefont
  {Wolf}(1986)}]{Sander1986Solubility}%
  \BibitemOpen
  \bibfield  {author} {\bibinfo {author} {\bibfnamefont {U.}~\bibnamefont
  {Sander}}\ and\ \bibinfo {author} {\bibfnamefont {B.~A.}\ \bibnamefont
  {Wolf}},\ }\href@noop {} {\bibfield  {journal} {\bibinfo  {journal} {Angew.
  Makromol. Chem.}\ }\textbf {\bibinfo {volume} {139}},\ \bibinfo {pages} {149}
  (\bibinfo {year} {1986})}\BibitemShut {NoStop}%
\bibitem [{\citenamefont {Fisher}\ and\ \citenamefont
  {de~Gennes}(1978)}]{Fisher1978Wall}%
  \BibitemOpen
  \bibfield  {author} {\bibinfo {author} {\bibfnamefont {M.}~\bibnamefont
  {Fisher}}\ and\ \bibinfo {author} {\bibfnamefont {P.~G.}\ \bibnamefont
  {de~Gennes}},\ }\href@noop {} {\bibfield  {journal} {\bibinfo  {journal} {C.
  R. Acad. Sci. Paris B}\ }\textbf {\bibinfo {volume} {287}},\ \bibinfo {pages}
  {207} (\bibinfo {year} {1978})}\BibitemShut {NoStop}%
\bibitem [{\citenamefont {Krech}(1994)}]{Krech1994The}%
  \BibitemOpen
  \bibfield  {author} {\bibinfo {author} {\bibfnamefont {M.}~\bibnamefont
  {Krech}},\ }\href@noop {} {\emph {\bibinfo {title} {The Casimir Effect in
  Critical Systems}}}\ (\bibinfo  {publisher} {World Scientific, Singapore},\
  \bibinfo {year} {1994})\BibitemShut {NoStop}%
\bibitem [{\citenamefont {Krech}(1997)}]{krech1997casimir}%
  \BibitemOpen
  \bibfield  {author} {\bibinfo {author} {\bibfnamefont {M.}~\bibnamefont
  {Krech}},\ }\href@noop {} {\bibfield  {journal} {\bibinfo  {journal}
  {Physical Review E}\ }\textbf {\bibinfo {volume} {56}},\ \bibinfo {pages}
  {1642} (\bibinfo {year} {1997})}\BibitemShut {NoStop}%
\bibitem [{\citenamefont {Krech}(1999)}]{krech1999fluctuation}%
  \BibitemOpen
  \bibfield  {author} {\bibinfo {author} {\bibfnamefont {M.}~\bibnamefont
  {Krech}},\ }\href@noop {} {\bibfield  {journal} {\bibinfo  {journal} {Journal
  of Physics: Condensed Matter}\ }\textbf {\bibinfo {volume} {11}},\ \bibinfo
  {pages} {R391} (\bibinfo {year} {1999})}\BibitemShut {NoStop}%
\bibitem [{\citenamefont {Hertlein}\ and\ \citenamefont
  {Helden}(2008)}]{hertlein2008gambassi}%
  \BibitemOpen
  \bibfield  {author} {\bibinfo {author} {\bibfnamefont {C.}~\bibnamefont
  {Hertlein}}\ and\ \bibinfo {author} {\bibfnamefont {L.}~\bibnamefont
  {Helden}},\ }\href@noop {} {\bibfield  {journal} {\bibinfo  {journal}
  {Nature}\ }\textbf {\bibinfo {volume} {451}},\ \bibinfo {pages} {172}
  (\bibinfo {year} {2008})}\BibitemShut {NoStop}%
\bibitem [{\citenamefont {Gambassi}\ \emph {et~al.}(2009)\citenamefont
  {Gambassi}, \citenamefont {Macio{\l}ek}, \citenamefont {Hertlein},
  \citenamefont {Nellen}, \citenamefont {Helden}, \citenamefont {Bechinger},\
  and\ \citenamefont {Dietrich}}]{gambassi2009critical}%
  \BibitemOpen
  \bibfield  {author} {\bibinfo {author} {\bibfnamefont {A.}~\bibnamefont
  {Gambassi}}, \bibinfo {author} {\bibfnamefont {A.}~\bibnamefont
  {Macio{\l}ek}}, \bibinfo {author} {\bibfnamefont {C.}~\bibnamefont
  {Hertlein}}, \bibinfo {author} {\bibfnamefont {U.}~\bibnamefont {Nellen}},
  \bibinfo {author} {\bibfnamefont {L.}~\bibnamefont {Helden}}, \bibinfo
  {author} {\bibfnamefont {C.}~\bibnamefont {Bechinger}}, \ and\ \bibinfo
  {author} {\bibfnamefont {S.}~\bibnamefont {Dietrich}},\ }\href@noop {}
  {\bibfield  {journal} {\bibinfo  {journal} {Phys. Rev. E}\ }\textbf {\bibinfo
  {volume} {80}},\ \bibinfo {pages} {061143} (\bibinfo {year}
  {2009})}\BibitemShut {NoStop}%
\bibitem [{\citenamefont {Rafa{\"\i}}, \citenamefont {Bonn},\ and\
  \citenamefont {Meunier}(2007)}]{rafai2007repulsive}%
  \BibitemOpen
  \bibfield  {author} {\bibinfo {author} {\bibfnamefont {S.}~\bibnamefont
  {Rafa{\"\i}}}, \bibinfo {author} {\bibfnamefont {D.}~\bibnamefont {Bonn}}, \
  and\ \bibinfo {author} {\bibfnamefont {J.}~\bibnamefont {Meunier}},\
  }\href@noop {} {\bibfield  {journal} {\bibinfo  {journal} {Phys. A}\ }\textbf
  {\bibinfo {volume} {386}},\ \bibinfo {pages} {31} (\bibinfo {year}
  {2007})}\BibitemShut {NoStop}%
\bibitem [{\citenamefont {Sengers}\ and\ \citenamefont
  {Shanks}(2009)}]{sengers2009experimental}%
  \BibitemOpen
  \bibfield  {author} {\bibinfo {author} {\bibfnamefont {J.~V.}\ \bibnamefont
  {Sengers}}\ and\ \bibinfo {author} {\bibfnamefont {J.~G.}\ \bibnamefont
  {Shanks}},\ }\href@noop {} {\bibfield  {journal} {\bibinfo  {journal} {J.
  Stat. Phys.}\ }\textbf {\bibinfo {volume} {137}},\ \bibinfo {pages} {857}
  (\bibinfo {year} {2009})}\BibitemShut {NoStop}%
\bibitem [{\citenamefont {Fukuto}, \citenamefont {Yano},\ and\ \citenamefont
  {Pershan}(2005)}]{Fukuto2005Critical}%
  \BibitemOpen
  \bibfield  {author} {\bibinfo {author} {\bibfnamefont {M.}~\bibnamefont
  {Fukuto}}, \bibinfo {author} {\bibfnamefont {Y.~F.}\ \bibnamefont {Yano}}, \
  and\ \bibinfo {author} {\bibfnamefont {P.~S.}\ \bibnamefont {Pershan}},\
  }\href@noop {} {\bibfield  {journal} {\bibinfo  {journal} {Phys. Rev. Lett.}\
  }\textbf {\bibinfo {volume} {94}},\ \bibinfo {pages} {135702} (\bibinfo
  {year} {2005})}\BibitemShut {NoStop}%
\bibitem [{\citenamefont {Vasilyev}\ \emph {et~al.}(2009)\citenamefont
  {Vasilyev}, \citenamefont {Gambassi}, \citenamefont {Macio{\l}ek},\ and\
  \citenamefont {Dietrich}}]{vasilyev2009universal}%
  \BibitemOpen
  \bibfield  {author} {\bibinfo {author} {\bibfnamefont {O.}~\bibnamefont
  {Vasilyev}}, \bibinfo {author} {\bibfnamefont {A.}~\bibnamefont {Gambassi}},
  \bibinfo {author} {\bibfnamefont {A.}~\bibnamefont {Macio{\l}ek}}, \ and\
  \bibinfo {author} {\bibfnamefont {S.}~\bibnamefont {Dietrich}},\ }\href@noop
  {} {\bibfield  {journal} {\bibinfo  {journal} {Phys. Rev. E}\ }\textbf
  {\bibinfo {volume} {79}},\ \bibinfo {pages} {041142} (\bibinfo {year}
  {2009})}\BibitemShut {NoStop}%
\bibitem [{\citenamefont {Hasenbusch}(2010)}]{hasenbusch2010thermodynamic}%
  \BibitemOpen
  \bibfield  {author} {\bibinfo {author} {\bibfnamefont {M.}~\bibnamefont
  {Hasenbusch}},\ }\href@noop {} {\bibfield  {journal} {\bibinfo  {journal}
  {Phys. Rev. B}\ }\textbf {\bibinfo {volume} {82}},\ \bibinfo {pages} {104425}
  (\bibinfo {year} {2010})}\BibitemShut {NoStop}%
\end{thebibliography}%

\end{document}